%% file: marin.tex
\documentclass[graybox]{svmult}

\usepackage{mathptmx}       % selects Times Roman as basic font
\usepackage{helvet}         % selects Helvetica as sans-serif font
\usepackage{courier}        % selects Courier as typewriter font
\usepackage{type1cm}        % activate if the above 3 fonts are
\usepackage{makeidx}         % allows index generation
\usepackage{graphicx}        % standard LaTeX graphics tool
\usepackage{multicol}        % used for the two-column index
\usepackage[bottom]{footmisc}% places footnotes at page bottom

\makeindex             % used for the subject index
\begin{document}

\title*{The panchromatic polarization signatures of Active Galactic Nuclei}
\author{Fr\'ed\'eric Marin}
\institute{F. Marin \at Universit\'e de Strasbourg, CNRS, Observatoire Astronomique de Strasbourg, UMR 7550, F-67000 Strasbourg, France, \email{frederic.marin@astro.unistra.fr}}
\maketitle

\abstract*{Among all the astronomical sources investigated through the prism of polarimetry, active galactic nuclei (AGN) have proven to be the richest in terms of 
complex yet fundamental signatures that helped to understand their true nature. Indeed, AGN exhibit a wide range of wavelength-dependent polarimetric features that 
are intrinsically related to their multi-scale emission mechanisms. Each waveband is characterized by a different set of polarimetric signatures that can be related
to various physical mechanisms that, from the radio band to the soft-$\gamma$ rays, probe increasingly smaller AGN regions. In fact, panchromatic polarization measurements
are the key to understand how and when AGN form, accrete, and impact the host galaxy they reside in. In this chapter, I will first introduce AGN without focusing on
a particular observational technique. I will then review the discoveries and constraints that spectro-, imaging and broadband polarimetry have achieved. Finally, I 
will highlight the important questions that remain unanswered and how they can be solved with future large millimeter and radio antennas, 30-m class optical and 
infrared telescopes, and high energy satellites equipped with state-of-the-art polarimeters.}

\abstract{Among all the astronomical sources investigated through the prism of polarimetry, active galactic nuclei (AGN) have proven to be the richest in terms of 
complex yet fundamental signatures that helped to understand their true nature. Indeed, AGN exhibit a wide range of wavelength-dependent polarimetric features that 
are intrinsically related to their multi-scale emission mechanisms. Each waveband is characterized by a different set of polarimetric signatures that can be related
to various physical mechanisms that, from the radio band to the soft-$\gamma$ rays, probe increasingly smaller AGN regions. In fact, panchromatic polarization measurements
are the key to understand how and when AGN form, accrete, and impact the host galaxy they reside in. In this chapter, I will first introduce AGN without focusing on
a particular observational technique. I will then review the discoveries and constraints that spectro-, imaging and broadband polarimetry have achieved. Finally, I 
will highlight the important questions that remain unanswered and how they can be solved with future large millimeter and radio antennas, 30-m class optical and 
infrared telescopes, and high energy satellites equipped with state-of-the-art polarimeters.}

%%%%%%%%%%%%%%%%%%%%%%%%%%%%%%%%%%%%%%%%%%%%%%%%%%%%%%%%%%%%%%%%%%%%%%%%%%%%%%%%%%%%%%
%%%%%%%%%%%%%%%%%%%%%%%%%%%%%%%%%%%%%%%%%%%%%%%%%%%%%%%%%%%%%%%%%%%%%%%%%%%%%%%%%%%%%%
\section{A brief introduction to AGN}
\label{AGN}
\begin{figure}[t]
  \centering
  \includegraphics[scale=0.5]{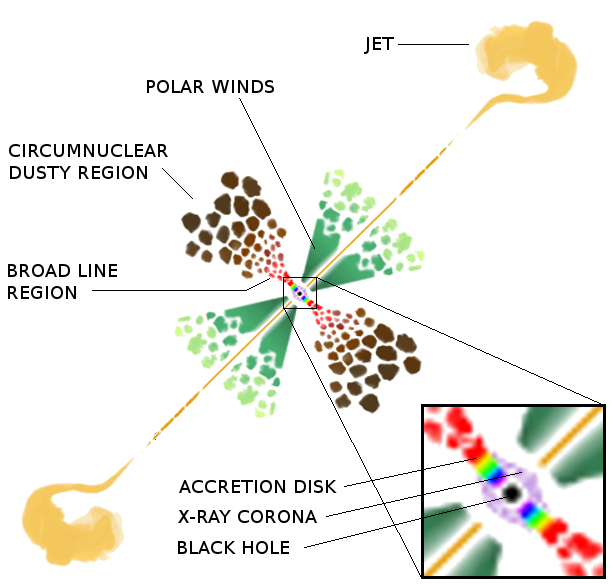}
  \caption{Unscaled sketch of an AGN. At the center lies a supermassive black hole 
	    around which a multi-temperature accretion disc spirals (shown with the 
	    color pattern of a rainbow). The accreted matter is illuminated by an 
	    X-ray corona (shown in violet) of unknown size. The disk extends from 
	    $\sim$ 10$^{-5}$ pc to $\sim$ 10$^{-3}$ pc for a 10$^8$ solar masses 
	    black hole. The region responsible for the emission of broad lines (BLR) 
	    is in red and light brown. It extends up to $\sim$ 10$^{-1}$ pc, where 
	    the circumnuclear dusty region (shown in dark brown) onsets. The
	    collimated polar ionized winds (in green) are created in sub-parsec 
	    scale regions and their final extension interacts with the interstellar 
	    medium (shown in yellow-green at a few hundreds of parsecs). A 
	    double-sided, kilo-parsec jet (in yellow) is added to account for 
	    radio-loud AGNs \cite{Marin2016b}.}
  \label{Fig:AGN} 
\end{figure}

``Active galactic nuclei'' (AGN) is the most general term used to describe a compact region at the center of a galaxy that often outshines the starlight contribution from 
the host, with characteristics indicating that the radiation is not produced by stars. With luminosities ranging from $\sim$ 10$^{40}$ to 10$^{47}$~erg.s$^{-1}$ for 
the most distant ones \cite{Mortlock2011}, AGN are considered as the most powerful, long-lived objects in the Universe. To explain the production of such tremendous 
amounts of radiation, accretion of gas and dust onto a supermassive black hole (SMBH) is invoked \cite{Pringle1972,Shakura1973}. The dissipative processes in the viscous 
accretion disc that forms around the SMBH transport matter inwards and angular momentum outwards, causing the accretion disc to heat up. The resulting thermal 
multi-temperature black body emission (see Fig.~\ref{Fig:AGN}) peaks in the ultraviolet and shapes the spectral energy distribution (SED) of AGN in the form of the 
\textit{Big Blue Bump} \cite{Malkan1982,Sanders1989}. The structure of the disk is still debated because of its spatially unresolvable sizes (smaller than a milliparsec 
for a 10$^8$ solar masses black hole). Depending on the dimensionless mass accretion rate $\dot{m}$, normalized to the Eddington rate of the central object, the accretion 
flow could sustain various geometries and natures \cite{Frank2002}. For $\dot{m} \le$ 1, the disk is likely geometrically thin, optically thick, and radiatively 
efficient \cite{Shakura1973}. For $\dot{m} >$ 1, the disk becomes geometrically slim (not thin), stays optically thick and is radiatively inefficient \cite{Abramowicz1988}.
However, if $\dot{m} \le$ 0.01, the accreted gas density is low so the gas may be unable to radiate energy at a rate that balances viscous heating. The flow becomes optically
thin and radiatively inefficient \cite{Narayan1994}. In all cases, it is commonly admitted that the inner edge of the accretion flow is set by the innermost stable circular 
orbit (ISCO) radius \cite{Misner1973}. Outside the ISCO, particles can orbit indefinitely in stable circular orbits while inside the ISCO they spiral rapidly past the 
event horizon into the SMBH. The location of the ISCO depends on the angular momentum (spin) of the central object. For a non-spinning massive object, where the 
gravitational field can be expressed with the Schwarzschild metric, the ISCO is located at 3$R_{\rm S}$ (Schwarzschild radius $R_{\rm S}$ = $\frac{2\,GM}{c^2}$). In 
the case of rotating black holes, the ISCO in the Kerr metric depends on whether the orbit is prograde or retrograde, resulting in a sign difference for the spin 
parameter. For massive particles around a maximally spinning black hole, the ISCO is at 0.5$R_{\rm S}$ (prograde), and at 4.5$R_{\rm S}$ for retrograde orbits. 

~\

Nearby the inner edge of the accretion disk should lie the region responsible for the emission of X-ray photons. Once again, its nature and geometry are poorly constrained, 
but it is believed that a plasma of hot electrons is situated on top of the disk, in an atmosphere-like region \cite{Haardt1993, Walter1993,Magdziarz1998} or in a compact 
or patchy corona situated at an unknown distance above the disk (\cite{Martocchia1996,Miniutti2004}, see the zoomed region in Fig.~\ref{Fig:AGN}). The corona may be generated 
by buoyancy instabilities in the disk and heated by magnetic reconnection, i.e. shocks at reconnection sites where strong impulsive heating occurs when magnetic field lines
are brought together \cite{DiMatteo1998}. The temperature of the plasma is such that ultraviolet photons, thermally emitted by the disk, are boosted up to the X-ray energies
by multiple inverse-Compton scattering. This results in a source of X-rays that can be variable in emission, such as observed with X-ray satellites \cite{Lawrence1985,McHardy1987}.
The power-density spectrum shows a power-law form with a turnover at low frequencies and a high frequency break, similar to galactic black-hole candidates (see the chapter
by Tariq Shabaz). Part of the X-ray photons that are emitted by the corona are bent back to the disk by general relativistic effects. Soft X-rays are mainly absorbed by
photoelectric absorption followed by fluorescent line emission, or photoelectric absorption followed by Auger de-excitation. Hard X-ray photons mainly scatter on the disk
due to Compton scattering by free or bound electrons \cite{George1991}. The most prominent signature of these physical effects is the presence of a strong iron fluorescence 
line at approximately 6.4~keV \cite{Nandra1994}. The emission line profile is often broad, skewed towards the soft X-rays by Doppler shifts and relativistic boosting due 
to the motion of the disk and the gravitational redshifting of the black hole \cite{Fabian2000}. Iron line fitting then allows to determine the radius of the ISCO, hence
the dimensionless spin parameter and the inclination of the accretion disk.

~\

The maximal extension of the outer rim of the accretion disk is, however, unconstrained. It is thought to coincide with the innermost parts of another fundamental constituents 
of AGN: the broad line region (BLR). Broad emission lines in the optical and ultraviolet bands are a key signature of AGN. They have been detected more than 50 years ago 
\cite{Sandage1965} and exhibit a variety of Doppler widths, from 1\,000~km.s$^{-1}$ to velocities larger than 15\,000~km.s$^{-1}$ \cite{Strateva2003}. Line emission results
from photoionization of the BLR gas by the thermal emission from accretion disk, with the BLR reaching a photoionization equilibrium temperature of 10\,000 -- 20\,000~K. 
Different gas densities, input ionizing spectra, gas dynamics, and elemental abundances impact the broad line profiles. Temperatures of the order of 10$^4$~K correspond
to thermal line widths around 10~km.s$^{-1}$, while different elemental abundances play a minor role in line broadening. It is then clear that the gas in the BLR must be 
dense ($n_{\rm e} >$ 10$^9$ cm$^{-3}$) and move supersonically in order to broaden the lines up to several thousands of kilometers per second \cite{Peterson2006}. But what 
is the geometry of this region? An uniformly filled, rotating BLR would not be able to produce the diversity of line emissions, such as the He{\sc ii}, N{\sc v} and C{\sc iv} 
high-ionization lines or the Mg{\sc ii}, Ca{\sc ii} and Fe{\sc ii} low-ionization lines. A distribution of discrete clouds, with different densities and distances from the 
center, is able to reproduce the total spectra of AGN \cite{Baldwin1995}, but its geometrical arrangement cannot be spherical otherwise we would observe Lyman continuum 
absorption by the BLR clouds. This, in fact, was never convincingly observed \cite{Antonucci1989,Gaskell2009}, suggesting that the BLR cloud distribution is likely flattened 
along the equatorial plane (often considered as coplanar to the accretion disk plane, see the light brown region in Fig.~\ref{Fig:AGN}). The resulting radial ionization 
stratification of the BLR, due to cloud self-shielding, allows the emission of both low and high ionization lines \cite{Gaskell2009}. The same idea was used by different 
models to explain the physical cause of the formation BLR, either from accretion disk gravitational instabilities \cite{Collin2001}, a failed radiatively accelerated dusty 
wind \cite{Czerny2017} or a fully formed, radiation-pressure-driven disk outflow that is self-shielded from the ionizing continuum \cite{Elvis2017}. The question of where 
does the disk ends and where does the BLR onset is yet to be answered but what is sure is 1) there is a tight observational relation between the BLR radius and the AGN 
bolometric luminosity \cite{Kaspi2000} and 2) the BLR most probably ends where dust grains can start to survive the intense ultraviolet radiation field \cite{Netzer1993}. 

~\

From the dust sublimation radius and beyond \cite{Barvainis1987,Kishimoto2007}, dust grains are though to form an optically thick, equatorial region historically coined with
the term ``torus'' \cite{Antonucci1993}. This is a necessary AGN constituent to explain the disappearance of the broad emission lines from the spectra of certain objects
\cite{Osterbrock1984}. An asymmetric dusty obscurer in the form of a donut prevents the detection of the BLR and accretion disk if the observer's viewing angle is crossing 
the compact dusty layers. Viewed from the top, the hole in the pastry allows the observer to have a direct view of the central engine. This lead to the still used denomination
of type-1 (dust-free, polar view) and type-2 (dust-obscured, equatorial view) AGN terminology. The donut picture evolved thanks to observational and numerical constraints 
\cite{Bianchi2012}, and our actual representation of the torus is a complex, clumpy medium that is in dynamical motion \cite{Garcia2016} (see Fig.~\ref{Fig:AGN}). From a 
type-1 view, once could see in principle that the inner part of this circumnuclear region is heated up to 1\,500~K by the accretion disk emission and cools mainly by thermal 
dust infrared emission. This phenomenon naturally produces the ubiquitous near-infrared bump/excess observed in the spectral energy distribution of AGN \cite{Neugebauer1979,Robson1986,Sanders1989}. 
The inner radius of this equatorial region is set by the luminosity of the photoionizing radiation originating in the central accretion disk \cite{Lawrence1991}. For ultraviolet 
luminosities between 10$^{43}$ and 10$^{44}$~erg.s$^{-1}$, the inner torus radius is typically comprised between 0.01 and 0.1~pc \cite{Marin2016a}. The maximal extension of 
this opaque dusty region was first though to expand as large as 100~pc \cite{Jaffe1993}, which is challenged by dynamical stability arguments, but the compact core of 
the torus is now observationally constrained to be only a dozen parsecs wide \cite{Jaffe2004}. A clumpy distribution of graphite grains embedded in a non-spherical
geometry can easily reproduce the 2 -- 4~$\mu$m excess for temperatures close to the dust sublimation limit \cite{Barvainis1987}. With increasing distances from the torus 
inner edges, the outer shells present colder dusty grains, only heated by reprocessed infrared radiation with characteristic temperatures lower than 100~K. The resulting 
infrared emission peaks at near- to mid-infrared wavelengths (3 -- 40~$\mu$m, \cite{Fritz2006}). Silicate grains are responsible for the observed absorption feature near 
9.7~$\mu$m we observe in dust-obscured AGN, while graphite grains are responsible for the rapid decline of the emission at wavelength shortwards of a few microns. Infrared 
emission from the torus should then decrease in the far-infrared/millimeter domain but observations often indicate a peak of emission between 50 and 100~$\mu$m 
\cite{Rowan-Robinson1984}. This is due to the contribution of starburst regions, in which ultraviolet radiation produced by the obscured star-formation are absorbed by the
enveloping dust layers and re-radiated around 100~$\mu$m. Strong correlations have been found between star formation activity and presence of polycyclic aromatic hydrocarbons
in the torus \cite{Woo2012}, demonstrating the importance of the host galaxy in the energy budget of the AGN SED. If large scale magnetic field are indeed present within 
the torus \cite{Lopez2013}, magnetically-aligned dust grains could produce polarized infrared emission and rotating grains could result in polarized radio continuum emission.
This make the circumnuclear dusty region of AGN an important part of the global puzzle.

~\

An interesting question is: what constrains the opening angle of the torus? AGN are very often associated with extended polar outflows that carry mass and energy from 
the sub-parsec central engine to the interstellar/intergalactic medium \cite{Wilson1996}. The most observed geometrical form of such outflows is a double cone that lies
along the direction of the AGN radio axis (see the dark green region in Fig.~\ref{Fig:AGN}). The base of this bi-conical, extended region is photo-ionized by oppositely 
directed beams of radiation emitted from the accretion disk \cite{Macchetto1994}. The polar outflows have sharp, straight edges that could indicate prior collimation by 
the torus itself. Hence, there might be a direct correlation between the opening angle of the torus and of the winds below 100~pc \cite{Storchi2015}. As the radial distance 
from the central engine increases, dust from the interstellar medium starts to appears and mix with the electrons. This slow-moving ($<$ 1\,000~km.s$^{-1}$), low-density 
(10$^3$ $\le n_{\rm e} \le$ 10$^6$~cm$^{-3}$) wind produces multiple narrow emission lines such as H{\sc i}, He{\sc i}, He{\sc ii}, [O{\sc iii}]$\lambda$4959, 
[O{\sc iii}]$\lambda$5007, [N{\sc ii}]$\lambda$6548 or [N{\sc ii}]$\lambda$6583. This narrow line region (NLR) is not uniform; it shows evidence for gas clouds, emission 
filaments, arcs-like structures, and a large gradient of the line-of-sight gas velocities, with absorption features uniformly blue-shifted relative to the systemic velocity
\cite{Afanasiev2007a,Afanasiev2007b}. This is indicative of a variety of processes happening in the NLR that are ultimately responsible for the presence of warm absorber-emitters
\cite{Porquet2000} or the detection of ultra-fast outflows close to the wind launching site \cite{Tombesi2010}. To explain the presence of ultra-fast outflows and the 
various NLR components we observe, several scenarios have been investigated and it was found that essentially three mechanisms could participate in launching a wind: thermal
driving, radiation pressure or magnetic forces. Accretion disks in hydrostatic equilibrium can produce a thermal wind if the cooler, outer part of the disk is irradiated by 
its hotter, inner regions. This would result in a high temperature at the disk surface that would puff up the upper gas layers in the forms of a static corona or an outflow 
\cite{Begelman1983a,Begelman1983b}. Another formation mechanisms relies on radiation pressure by electron scattering (if the wind is fully ionized) or by line scattering. 
Spectral lines increase the scattering coefficient, giving rise to powerful line-driven winds \cite{Proga1998,Proga1999,Higginbottom2014}. According to the authors, an 
accretion disk around a 10$^8$ solar masses SMBH accreting at the rate of 1.8 solar masses per year can launch a wind at 0.003~pc from the central engine, reaching velocities 
up to 0.05$c$ (at 0.03~pc from the potential well). Finally, massive turbulent strongly magnetized thick gas disk can give rise to magnetocentrifugal winds \cite{Vollmer2018}. 
In fact, the magnetorotational instability is an universal mechanism to produce turbulence and transport angular momentum in disks at all radii \cite{Balbus1998}. 

~\

Strongly magnetized accretion disks are not only responsible for bulk outflows; they can produce highly collimated, powerful jet that can reach to the intergalactic medium.
The most impressive, largest jets can reach up to a few mega-parsecs in projected linear sizes \cite{Riley1989,Bhatnagar1998,Schoenmakers2001}, while more regular jets 
easily reach dozens of kilo-parsecs. Unscaled jets are presented in yellow in Fig.~\ref{Fig:AGN}, with narrow bases and with two lobes of radio emission more or less 
symmetrically located on either side of the AGN. There are two main methods to extract enough power from the central AGN engine to form such jets: the Blandford-Znajek 
\cite{Blandford1977} and the Blandford-Payne \cite{Blandford1982} mechanisms. The Blandford-Znajek process requires an accretion disc with a strong poloidal magnetic field 
around a spinning black hole. Open magnetic field lines are transferring energy and angular momentum from the disk to the polar direction, resulting in Poynting flux dominated
jets. On the other hand, the Blandford-Payne process does not require a spinning black hole. The magnetic field threading the disk extracts energy from the rotating gas to 
power a jet within the co-rotating large scale magnetic fields. Which mechanism dominates is not yet determined \cite{Chai2012,Cao2018}. What is certain, however, is that jets 
are detected only in a small fraction of AGN. Depending on their ratio of 5~GHz to B-band flux density and radio power, about 15 -- 20\% of all AGN can be considered as radio-loud
\cite{Kellermann1989,Urry1995}. The difference between radio-loud and radio-quiet objects (in fact radio-faint, because there is some radio emission in radio-quiet AGN) echoes
over the whole electromagnetic spectrum \cite{Padovani2017}. Radio-loud AGN are dominated by non-thermal emission from their jet from the radio to the $\gamma$-ray band, while 
radio-quiet AGN are dominated by thermal emission. In addition, the radio-loud fraction of AGN is a function of redshift and optical and X-ray luminosities 
\cite{dellaCeca1994,LaFranca1994,Jiang2007}. It means that AGN, with their tremendous intrinsic brightness, should act as flashlights to illuminate the intergalactic medium 
at different cosmological periods, allowing us to study the conditions and transitions of the Universe early in cosmic history.

~\

As they propagate from the AGN up to galactic of intergalactic medium, jets deposit radiation and kinetic energy into the host galaxy. This feedback effect is responsible for
quenching starburst activities, creating red extended and dispersion-dominated galaxies. AGN activity in the form or jets, winds, or intense radiation can heat up the hydrogen
gas in the galaxy or blow it out completely, thus preventing the gas from cooling and contracting to form stars \cite{Bower2006,Dubois2013}. Thus, there is a direct action of 
the AGN on the galaxy it resides in, but the opposite may be also true. Both recent observations and simulations indicate that there is a delay of 50 -- 250~Myr between the onset 
of starburst and AGN activity \cite{DiMatteo2005,Davies2007}. This delay is due to a viscous time-lag as the gas from the host takes time to flow down to the AGN central engine
\cite{Blank2016}. There are increasing evidences for this mass accretion duty cycle in AGN, showing that the AGN-host system is tightly coupled. This coupling also impact any 
AGN observation: since the central engine is spatially unresolved, starlight and dust emission from the host often contaminates the AGN signal. This is particularly true in the 
optical, near-infrared and far-infrared domains, where parasitic light often dominates in the case of lesser luminous AGN (low-luminosity AGN and Seyfert-galaxies). It results 
that any complete AGN picture cannot be dissociated from its host galaxy, a statement that will reach its highest significance when we will discuss polarimetry.

%%%%%%%%%%%%%%%%%%%%%%%%%%%%%%%%%%%%%%%%%%%%%%%%%%%%%%%%%%%%%%%%%%%%%%%%%%%%%%%%%%%%%%
%%%%%%%%%%%%%%%%%%%%%%%%%%%%%%%%%%%%%%%%%%%%%%%%%%%%%%%%%%%%%%%%%%%%%%%%%%%%%%%%%%%%%%
\section{Successes of polarimetric observations in constraining the AGN physics}
\label{History}
The global picture of AGNs seems rather well understood. In fact, details about the accretion physics, formation mechanisms, magnetic fields, AGN constituent morphologies, 
and kinetic and radiative interactions between the various AGN components are very elusive. The main problem we have to cope with is the cosmological distance of AGNs. Among 
the nearest type-1 AGNs is NGC~3227 (10h23m30.5790s, +19d51m54.180s), situated at an heliocentric redshift of 0.00386 $\pm$ 0.00001 (H$_{\rm 0}$ = 73 km.s$^{-1}$.Mpc$^{-1}$, 
$\Omega_{\rm matter}$ = 0.27, $\Omega_{\rm vacuum}$ = 0.73). This redshift-distance corresponds to an Hubble distance of 20.28 $\pm$ 1.45~Mpc. If we aim at resolving the 
ISCO of the accretion disk in NGC~3227, considering a central SMBH of 3.6~$\times$~10$^7$ solar masses \cite{Onken2003} and its associated 3.45~$\times$~10$^{-6}$~pc
Schwarzschild radius, about 35~nano-arcsec spatial resolution would be required. This is beyond any prospects, even for X-ray interferometry \cite{Skinner2009}. Since 
regular observational techniques such as photometry, spectroscopy or timing analyses have proven to be not sufficient to spatially resolve the AGN inner regions, a
work-around has been found: polarimetry. Polarization has the advantage of being wavelength-dependent and its main strength is that it is extremely sensitive to the geometry
and magnetic fields of the emitting/scattering region. Unlike spectroscopy that is limited by the physical size of the emitting region, polarization allows us to probe 
spatially unresolved volumes and still determine their geometry. Polarization is fundamentally linked to the internal properties of the sources of radiation: the strength 
and orientation of magnetic fields, the distribution and orientation of scattering particles like dust grains, the microscopic structure of reflecting surfaces, or intrinsic
anisotropies of the host galaxy surrounding AGNs. Hence, by measuring the polarization of AGNs in different wavebands of the electromagnetic spectrum it is possible to 
constrain the geometry of the innermost, spatially unresolvable, components. In the following I will review the greatest successes of polarimetric observations in constraining 
the AGN physics, complementing the work of previous authors \cite{Antonucci2002}. The list of discoveries is roughly in the chronological order. I will not review in details 
the mechanisms of polarization production as they are covered in a dedicated chapter of this book.

\subsection{The origin of the ``featureless'' continuum}
\label{History:continuum}
The origin of the optical and ultraviolet continuum emission in AGNs has been long discussed in the past \cite{Schmidt1980,Serote1996}. The general form of the white light
SED is a power-law F$_\lambda \sim \lambda^\alpha$, with $\alpha$ the spectral photon index of the order of 1 -- 3 \cite{Page2005,Corral2011}. The origin of this power-law
emission was for long puzzling as both synchrotron emission or thermal emission from a multiple-temperature black body (the accretion disk) could explain the spectral shape. 
In addition, continuum photons and absorption lines from the stellar atmospheres of old stellar populations within the host galaxy tend to complicate the problem by diluting
the signal \cite{Kruszewski1971}. This is especially true in type-2 AGNs where the host dominates over nuclear emission. 

Spectroscopy and imaging methods were, at the time (1970 -- 1980), not able to solve the origin of the ``featureless'' continuum (equivalent width of emission lines $<$~5~\AA). 
Polarization, on the other hand, was a key to elucidate this mystery. Indeed, if the measured continuum polarization comes from non-thermal emission by electrons in an anisotropic
magnetic field, the polarization is intrinsic. The polarization should then vary together with the total flux as the magnetic configuration is evolving, resulting in a polarized 
flux that should be proportional to the total flux at each wavelength \cite{Angel1980,Webb1993}. Synchrotron polarization is then expected to be strongly variable and different
to that of the polarization from the emission lines. On the other hand, if the polarization of the featureless continuum originates from scattering by electrons, dust grains or
atoms, the polarization should be insensitive to flux variations as the \textit{geometry} of the scattered cannot change over rapid (hours-days) timescales. The polarization 
of the continuum should be similar to that of the polarization in the emission lines if they originate from the same source. The scattering-induced polarization is also carrying 
wavelength-dependent signatures that can help to identify the nature of the scatterer. If Thomson scattering prevails, the polarization fraction is wavelength-independent. If 
scattering and transmission of light by a non-spherical distribution of dust grains is responsible for the polarization of the featureless continuum, the polarization fraction 
will vary as $\lambda^{-w}$ (with 1 $\le w \le$ 4 depending on the dust prescription \cite{Mathis1977}).

Using polarization measurements, it was then possible to conclude that the featureless continuum of 3C~84 (NGC~1275) is clearly related to synchrotron emission due to its 
month-to-month polarization variations, strengthening up by 3\% and rotating from 100 to 150$^\circ$ in polarization angle \cite{Walker1968,Babadzhanyants1975}. Later studies 
showed that the polarization of 3C~110 and 3C~246 is also attributable to non-thermal emission due to the rise of their polarized fluxes at longer wavelengths with respect to 
their total fluxes \cite{Webb1993}. In the same paper, the low levels of wavelength-independent, non-variable polarizations in NGC~4151, Mrk~509, NGC~5548 and Mrk~290 is better 
explained by thermal photons originating from and scattering in an accretion disk. Polarimetry was also used to separate the nuclear and galaxy components, the polarization of 
the former being diluted by the unpolarized starlight emission of the latter \cite{Visvanathan1968}. In fact, it was shown that in several sources such as NGC~7130, NGC~5135
and IC~3639 the nuclear starburst light dominates in the ultraviolet band, being directly responsible for the featureless continuum \cite{Gonzalez1998}. Polarimetry was thus
of invaluable help to better understand the complex and numerous emission processes that are shaping the SED of AGNs.

\subsection{The correlation between the AGN radio axis and optical polarization angle}
\label{History:Pol_angle}
While acquiring optical polarimetric measurements of a large sample of radio-loud AGNs with broad emission lines, it was discovered that their polarization position angle was 
strongly correlated with the direction of their jets \cite{Stockman1979}. Over the 24 initial targets, 20 had their optical polarization angle and radio position angle aligned 
within 30$^\circ$, and all presented low, $\le$~1~\%, polarization degrees. This correlation could no be of random nature but the physical mechanism of such alignment remained 
unknown at that time. Latter studies of radio-loud AGNs with and without broad emission lines showed that a given population (AGN with broad emission lines) displayed an alignment
between the position angle of optical polarization and the large scale radio structure, while a second population (AGN without broad emission lines) had a a perpendicular 
relationship \cite{Antonucci1982}. The same bi-modality was found in radio-quiet AGNs \cite{Antonucci1983,Martin1983} thanks to deep radio observations that discovered small
and weak radio structures despite the presence of jets. All sources exhibited weak polarization degrees, preventing a clear determination of the physical causes of such 
alignment. Scattering, transmission through dust grains or synchrotron radiation were suggested as possible answers but another key was necessary to unlock the secret.
Spectropolarimetry brought it a few years later.

\subsection{Unifying type-1 and type-2 AGNs}
\label{History:Unified_Model}

\begin{figure}[t]
    \centering
    \includegraphics[scale=1.5]{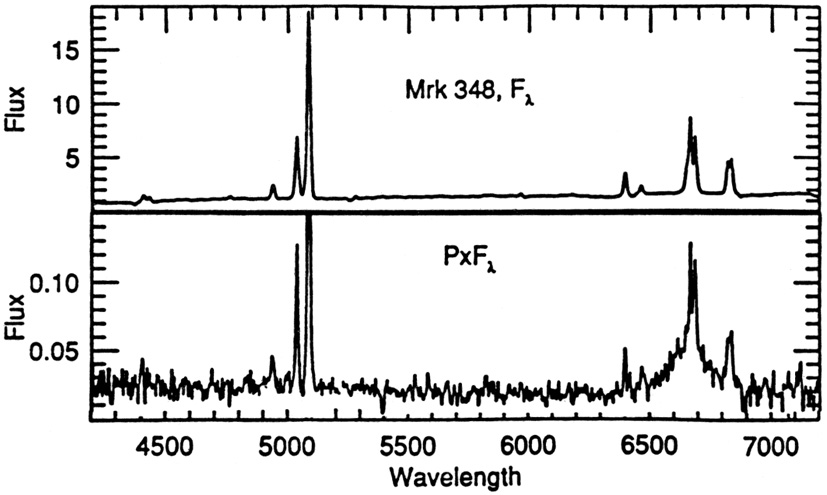}
    \caption{Total flux (top) and polarized flux (bottom)
	      spectra of the radio-quiet, type-2 AGN Mrk~348.
	      In polarized flux a very broad H$\alpha$ 
	      line can be seen. The relative flatness of 
	      the polarized continuum suggests that 
	      electron (Thomson) scattering is the prevalent 
	      the mechanism in Mrk~348 \cite{Miller1990}.}
    \label{Fig:Broad_line} 
\end{figure}

The puzzling question of the correlation between the AGN radio axis and optical polarization angle, and its bi-modality between AGN with and without optical broad emission 
lines, was investigated in further details using high signal-to-noise, high resolution spectropolarimetry. Among a small sample of radio-loud AGNs, 3C~234 stood out due to its 
high polarization degree (about 14\%) and its polarization position angle that is perpendicular to the radio axis \cite{Antonucci1984}. More importantly its continuum 
polarization, that is almost wavelength-independent, shares the same polarization properties than the broad H$\alpha$ line that is observed in polarized flux. The authors 
concluded that the polarization could be produced by a geometrically thick, optically thick electron scattering ring \cite{Antonucci1984}. The key was there but the lock 
squeaked. The final click was brought by \cite{Miller1983} and \cite{Antonucci1985} who discovered a highly polarized, $\sim$ 16~\%, wavelength-independent non-stellar 
continuum in NGC~1068. The polarized flux spectrum of this radio-quiet type-2 AGN revealed the presence of very broad (about 7500~km.s$^{-1}$) symmetric Balmer and 
permitted Fe{\sc ii} lines, that are very similar to what is observed in type-1 AGNs. The same polarization was measured in the polarized broad line and in the continuum, 
indicating a common origin. The perpendicular polarization angle and the high polarization degree of NGC~1068 became the most egregious evidence for the presence of an 
obscuring circumnuclear medium that is hiding the source of emission. Photons are scattering inside an electron-filled, polar region situated above and below the obscuring
disk, carrying broad line photons into the line-of-sight \cite{Antonucci1985}. This was the first compelling evidence for the presence of a type-1 nucleus inside a type-2 object. 
Searches for additional type-1 signatures in the polarized flux of other types-2 AGNs were successful \cite{Miller1990,Tran1992,Wills1992,Tran1995,Ogle1997}, see also 
Fig.~\ref{Fig:Broad_line}, proving that \textit{nuclear orientation} is one of the key parameters to understand the AGN zoology \cite{Antonucci1993}. 

Spectropolarimetry has shown that, for type-2 objects, broad lines are hidden behind an obscuring dusty layer which is now also observed in the radio and infrared thanks to adaptive
optics \cite{Jaffe2004} and interferometry \cite{Kishimoto2009}. However, it is impossible to prove that all type-2 AGNs have a broad emission line region. High-resolution spectra
and broadband observations (covering several Balmer lines) has proven to be necessary to observe faint broad lines in polarized flux \cite{Ramos2016}, decreasing the estimated 
number of true type-2s, i.e. AGN that are genuinely lacking a BLR. No strong constraints have been found between the absence of broad emission lines observed in polarized fluxes
and a given physical parameter such as bolometric luminosity, torus sizes or galaxy inclination \cite{Ramos2016}. However, the discovery of changing-look AGNs in which strong 
flux variations (up to 2 orders of magnitude) and the appearance/disappearance of broad emission lines may be a step to solve this issue \cite{Matt2003,LaMassa2015,Mathur2018}.
The disappearance of BLR signatures and the flux dimming could be explained by a variety of processes, from variable obscuration of the source to rapid mass accretion rate drop
\cite{Noda2018}. Polarimetry offers a natural way to investigate this question since any change in the geometry of the scatterer directly impact the polarization degree and angle
\cite{Marin2017}. In the case of the changing-look AGN J1011+5442 that switched between type-1 and type-2 classification between 2003 and 2015, the almost null polarization 
recorded in the type-2 phase indicates that the observed change of look is not due to a change of obscuration in the torus hiding the BLR and the central engine \cite{Hutsemekers2017}.
It is rather due to a rapid decrease of the SMBH accretion rate that is responsible for the vanishing of the BLR region \cite{Hutsemekers2017,Marin2017}. Additional polarimetric 
observations are necessary to determine the physical reasons behind the change of look of those AGNs, but this can only be done case-by-case.

\subsection{Revealing the hidden nuclear location}
\label{History:hidden_nucleus}

\begin{figure}[t]
    \sidecaption[t]
    \includegraphics[scale=.3]{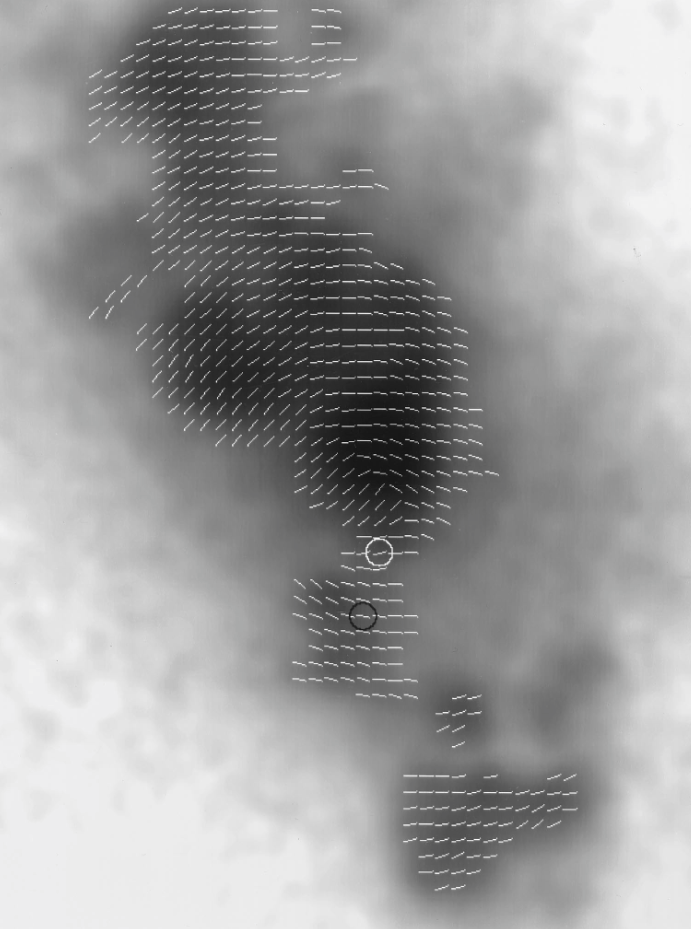}
    \caption{Hubble Space Telescope (HST) imaging polarimetry
	      of the inner region of the radio-quiet type-2 
	      AGN NGC~1068 taken with the COSTAR-corrected
	      Faint Object Camera in the 2400 -- 2700~\AA~
	      \cite{Capetti1995a}. The polarization vectors 
	      are superimposed to the total intensity map. 
	      The length of the vectors are not representative
	      of the polarization degree. The position of 
	      the source of scattered radiation is 
	      indicated using a white circle. The black
	      circle marks previous estimates.}
    \label{Fig:HST_map} 
\end{figure}

Closely following the foundation of the Unified Model of AGN \cite{Antonucci1993}, polarization mapping of the extended structures around the hidden core of type-2 AGNs 
allowed to pinpoint the source of emission. Back in 1990 -- 1995, the exact location of the central engine in dust-obscured sources was not easily determined (see Fig.~\ref{Fig:HST_map}).
Infrared \cite{Braatz1993} and radio \cite{Ulvestad1987} images were not in agreement regarding the question of the hidden nuclear location. For the closest objects, they diverged
by more or less one arcsecond, which represents several tens of parsecs in the case of NGC~1068 (which is among the most observed type-2s ever). Considering that the central engine
has a sub-parsec scale, such offset can be dramatic for the understanding of the internal physics of type-2 objects. Polarimetric attempts to determine the true location of the 
nucleus of NGC~1068, using scanning polarimeters, were done back in the 70's \cite{Elvius1978} but the polarization imaging capabilities brought by the Hubble Space Telescope (HST)
allowed to settle the debate. Using the Faint Object Camera (ultraviolet) and the Wide Field Planetary Camera (visual), with spatial resolutions of 0.06 and 0.08 arcseconds 
respectively, it was possible to map the extended NLR of nearby type-2 AGNs \cite{Capetti1995b,Capetti1995c}. The spatial resolution of the maps associated with the polarization
information allowed to reveal a double-conical shape with sharp edges, standing out with much more contrast than in total flux \cite{Macchetto1994}. The centro-symmetric pattern 
of the polarization vectors are pointing towards the true, hidden location of the central engine, such as illustrated in Fig.~\ref{Fig:HST_map}. The uncertainties in the source 
location is of the order of 0.02 arcseconds. A deeper look at the polarization structure in the NLR was achieved with repeated HST polarimetric mapping and allowed to isolate 30 
different clouds in the upper and lower NLR bicone \cite{Kishimoto1999}. The author used the clouds polarization degree as an indicator of the scattering angles to reconstruct 
the three-dimensional structure of the nuclear region. This lead to the first and, 20 years later, still unique three-dimensional view of an AGN that only polarimetry can achieve.

\subsection{The Laing-Garrington effect}
\label{History:LG_effect}

\begin{figure}[t]
    \sidecaption[t]
    \includegraphics[scale=.3]{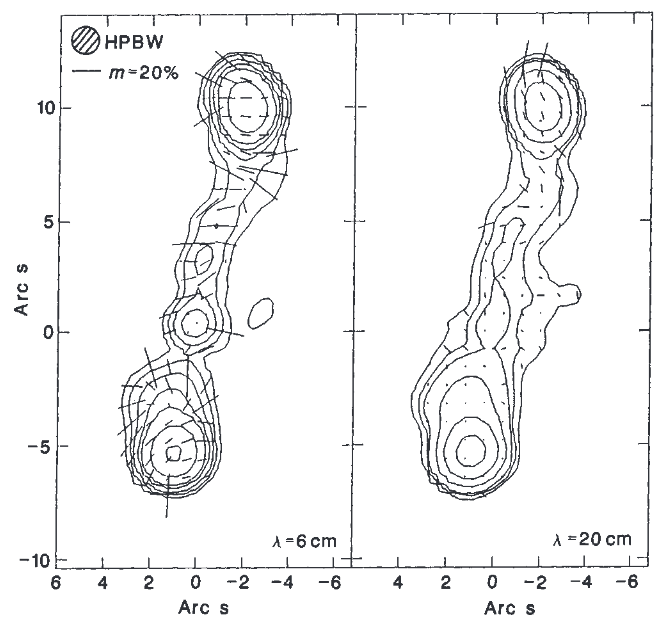}
    \caption{Radio maps of 4C16.49, a type-2 radio-loud 
	      AGN that shows prominent twin jets. The left 
	      map is taken at 6~cm while the right map is 
	      taken at 20~cm. The solid contours are 
	      proportional to the total intensity, while 
	      the vectors indicate the polarization angles.
	      The length of the vectors are proportional 
	      to the fractional polarization. The lobe 
	      containing the brighter jet depolarizes 
	      less rapidly with increasing wavelength 
	      than the lobe on the counter-jet side 
	      \cite{Laing1988,Garrington1988}.}
    \label{Fig:LG_effect} 
\end{figure}

Another very important piece of information about the physics of AGN has been brought by imaging polarization, this time in the radio band. Between the crucial observation
of NGC~1068 and the establishment of the Unified Model of AGNs, Very Large Array (VLA) observations shed light on the complex question of the one-sidedness in the jets of 
otherwise symmetrical extragalactic radio sources. If today we are almost certain that special relativistic effects (Doppler boosting) strongly favors relativistic jets 
and components approaching us, the absence of (detected) counter-jets was quite puzzling at that time (1970 -- 1990). Polarimetric mapping of the jets and lobes in a small 
sample of radio-loud AGNs have shown that, for double jetted AGN with one jet being brighter than the other, the lobe containing the brighter jet is almost always less depolarized 
with increasing wavelength than the lobe on the other counter-jet side \cite{Laing1988,Garrington1988}. This depolarization asymmetry, now known as the ``Laing-Garrington 
effect'' and visible in Fig.~\ref{Fig:LG_effect}, is of particular importance as it can be explained either by internal differences between the two lobes if the jet 
one-sidedness is intrinsic \cite{Garrington1988}, or by a differential Faraday rotation through irregularities in a magneto-ionic medium surrounding the radio source if 
the one-sidedness is due to Doppler beaming \cite{Laing1988}. In the second scenario, the side with the stronger jet closer to us is seen through a smaller amount of 
material and therefore shows less depolarization. This is intrinsically related to the orientation of the AGN, another major step that lead to the Unified Model a few
years later.

\subsection{The highly non-spherical structure of outflows}
\label{History:Outflows}
Outflows, together with jets, play a major role in the enrichment of the AGN host galaxy (see Sect.~\ref{History:Outflows}). Polar winds are known to exhibit a wide 
range of emission and absorption lines, among which broad absorption lines (BAL, only seen for the most luminous AGNs so far) have several fascinating properties. First, 
the BAL correspond to resonance lines of highly-ionized species such as C{\sc iv}, S{\sc iv}, or N{\sc v}. Second, the absorption features are blue-shifted with respect 
to the corresponding emission lines, with ejection velocity of the order of a fraction of the speed of light. Only 10 -- 20~\% of AGNs show such features \cite{Lamy2004} 
but the close similarity between the emission-line and continuum properties of BAL and non-BAL AGNs rules out a different nature. It is more likely that orientation plays
a role in the detection of BAL \cite{Weymann1991}. To decipher the origin of the outflows and the mechanisms responsible for their production, continuum, absorption and 
emission-lines polarization was scrupulously observed \cite{Hines1995,Brotherton1997,Hutsemekers1998b,Ogle1999,Lamy2004}. BAL AGNs appear to be more highly polarized than 
non-BAL objects, with polarization degrees neighboring 4~\% on average, which was first thought to be consistent with a more equatorial viewing direction \cite{Ogle1999}. 
However, their broad emission lines are often unpolarized \cite{Hines1995} and the polarization degree and position angle are wavelength-depend both in the continuum and 
absorption lines \cite{Brotherton1997}. This suggests multiple scattering mechanisms, resulting in a complex system geometry \cite{Lamy2004}. The increase of continuum 
polarization in the ultraviolet, associated with a regular rotation of the polarization position angle observed in [HB89]~0059-274 tend to point to two different polarization
mechanisms \cite{Lamy2000}. This is supported by polarization microlensing observations, where the microlensed polarized continuum comes from a compact region coplanar 
to the accretion disk while the non-microlensed continuum arises from an extended region located along the polar axis \cite{Hutsemekers2015}. The origin of the BAL wind 
might very well be related to an outflow originating from the disk or from the torus, and extending towards the polar axis. The continuum radiation would then be 
scattered inside the accretion disk and wind base, then inside the polar outflows, producing roughly perpendicular polarizations \cite{Hutsemekers1998b,Marin2013}.
This non-spherical structure is supported by the strongly polarized residual light within the broad absorption lines, implying that the outflows must have a small 
opening angle \cite{Ogle1999,Marin2013}. Spectropolarimetry has proven to be particularly insightful in this regard: the structures observed in polarized light 
across the broad H$\alpha$ emission line in PG~1700+518 indicate that the outflows (showing $\sim$ 4\,000~km.s$^{-1}$ rotational motions) must originate close to 
the accretion disk and rise nearly vertically \cite{Young2007}. The geometry and acceleration mechanisms of BAL winds is now actively discussed and computed as 
they appear to be a universal signature of massive and accreting objects \cite{Lucy1970,Proga2000,Elvis2000,Elvis2017}.

\subsection{Alignment of AGN polarization with large-scale structures}
\label{History:Large_scale_Structures}

\begin{figure}[t]
    \sidecaption[t]
    \includegraphics[scale=.32]{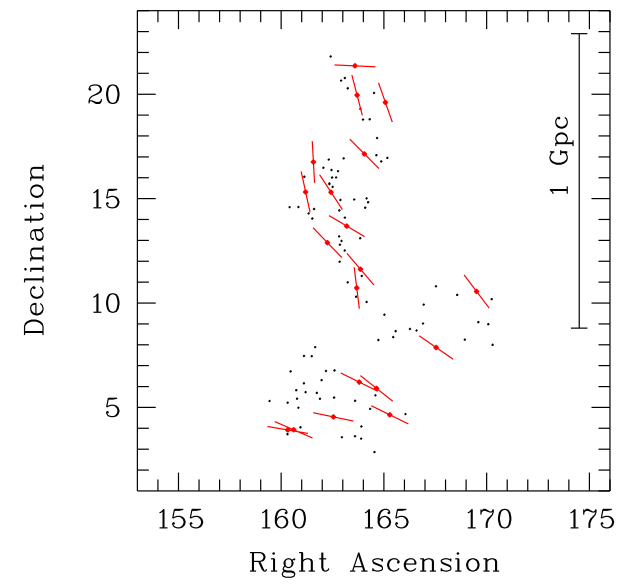}
    \caption{Polarization vectors (in red) of 19 radio-quiet
	      AGNs with a polarization degree larger than
	      interstellar contaminant polarization. The 
	      polarization vectors are superimposed on the 
	      large-scale structure belonging to Gpc scale 
	      AGN groups at redshift $z \approx$ 1.3 
	      \cite{Hutsemekers1998,Hutsemekers2001,Hutsemekers2014}.
	      Despite a clear correlation, the interpretation 
	      of this orientation effect remains puzzling.
	      A solution might be that the spin axes of AGNs 
	      are possibly parallel to their host large-scale 
	      structures.}
    \label{Fig:Quasar_alignement} 
\end{figure}

The orientation of AGNs is apparently not strongly correlated to the orientation of their host galaxy. At best, the extended radio structures of type-1 AGN are 
found to avoid alignment with their host galaxy major axis while type-2 AGN are more randomly distributed. In addition, both AGN types apparently avoid close 
alignment between their radio axis and their host galaxy plane axis \cite{Schmitt1997}. Apart from those specific zones of avoidance, there is no observational 
proofs nor reasons for a preferred direction of AGN polarization position angles. Yet, compiling the polarization position angle of 170 moderate-to-high redshift 
($z \sim$ 1 -- 2) AGNs, a concentration of polarization vectors with a preferential direction was found in a $\sim$ 1000~$h^{-1}$~Mp region of the sky 
\cite{Hutsemekers1998}. Subsequent measurements confirmed that the polarization vectors are coherently oriented in several groups of 20 -- 30 AGNs, which are 
roughly parallel to the plane of the Local Supercluster \cite{Hutsemekers2001}. However this result only holds at $z \ge$ 1, which rules out large scale 
($\sim$ 50~Mpc) magnetic fields that are either 1) converting photons into massless or extremely light pseudo-scalars \cite{Harari1992} or 2) being responsible 
for dichroic extinction and scattering \cite{Wood1997}. This means that the causes for AGN polarization orientation might very well be cosmological. 
Early large-scale primordial magnetic fields could have played a role on galaxy formation and orientation during the epoch of inflation, explaining the 
$z \ge$ 1 constraints \cite{Battaner1997,Hutsemekers2001}. More recent observations of radio-loud AGN at various redshifts have confirmed the alignment of 
AGN polarization with large-scale structures, such as shown in Fig.~\ref{Fig:Quasar_alignement} and detailed in \cite{Taylor2016,Contigiani2017}. Comparing 
the AGN optical polarization position angle to their host large-scale structures, it was additionally deduced that type-1 AGNs are preferentially perpendicular
to the host structure \cite{Hutsemekers2014}. This suggests that the spin axis of AGNs is parallel to their host large-scale structures.

\subsection{Accretion signatures in polarized fluxes}
\label{History:Disk_spectrum}

\begin{figure}[t]
    \sidecaption[t]
    \includegraphics[scale=.32]{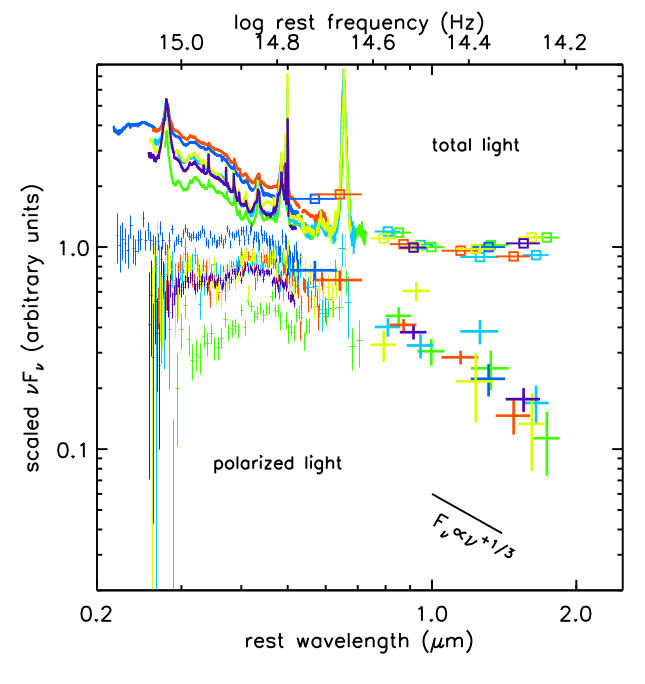}
    \caption{Overlay of the total and polarized flux spectra
	      of six different radio-loud, type-1 AGN \cite{Kishimoto2008}.
	      The polarized flux allows to shave off both 
	      the emission lines and the unpolarized re-emission 
	      by dust. All the objects behave in a similar and 
	      systematic way, revealing the expected blue 
	      polarized-light spectra of accretion disks.
	      This is a perfect example of how polarimetry 
	      can probe spatially unresolved regions, up 
	      to $\sim$ 800~$R_{\rm S}$ in this case.}
    \label{Fig:Disk_spectrum} 
\end{figure}

One of the most recent discoveries achieved with polarimetric measurements concerns the spatially non-resolvable inner regions of AGNs. The accretion disk 
that powers the SMBH is thought to radiate from the far-ultraviolet to the near-infrared \cite{Shakura1973}, but the disk near-infrared emission is 
diluted by quantitatively much larger dust emission from the bulk circumnuclear region (the torus). The spectral shape of the disk emission decreases 
rapidly at longer wavelength, up to 1 -- 2~$\mu$m, such as F$_\nu \sim \nu^{1/3}$. The addition of the re-emitted infrared light from the torus 
causes the spectral slope to be much redder \cite{Zheng1997,Vanden2001}. As a consequence, it is impossible to observe the total flux from the outer rim
of the disk, and thus confirm the disk paradigm \cite{Hubeny2000}. A fascinating aspect of polarimetry is that it produces a blueprint of the original 
spectrum and keeps memory of it as long as no additional scattering or absorption events happen. Looking at type-1 inclinations, the central engine is 
directly visible through the optically thin polar outflows. Light does not suffer from additional scattering other than in the equatorial plane, where 
the accretion disk is located. So, it is theoretically possible to observe the near-infrared component of the accretion disk in polarized light. Such piece 
of work was achieved in six different, local ($z \sim$ 0.2 -- 0.6) AGNs \cite{Kishimoto2009}. It was found that, as expected, the total flux spectra 
are dominated by hot dust re-emission around 1~$\mu$m, while the polarized flux spectra consistently and systematically decreasing towards the infrared,
see Fig.~\ref{Fig:Disk_spectrum}. The weighted mean of the spectral slope is 0.44 $\pm$ 0.11, which is very close to the theoretical limit of 0.33.
Interestingly, the authors do not find any correlation between the spectral slopes and the AGN black hole masses or Eddington luminosities. The marginal,
bluer spectral slope found could, however, be interpreted as a hint for a truncated (or a gravitationally unstable) disk. This opens the door for future
investigations of the continuum polarimetric signal in nearby AGNs.

%%%%%%%%%%%%%%%%%%%%%%%%%%%%%%%%%%%%%%%%%%%%%%%%%%%%%%%%%%%%%%%%%%%%%%%%%%%%%%%%%%%%%%
%%%%%%%%%%%%%%%%%%%%%%%%%%%%%%%%%%%%%%%%%%%%%%%%%%%%%%%%%%%%%%%%%%%%%%%%%%%%%%%%%%%%%%
\section{Current status on the panchromatic polarization of AGN}
\label{Status}
We have seen that polarimetric measurements, from the radio to the ultraviolet band, have lead to considerable discoveries. The most significant one is 
the establishment of the Unified Model of AGNs, but polarimetry has also proven to be able to achieve great discoveries from the scale of the accretion disk 
to the kilo-parsec jets. Each waveband is characterized by a different set of polarimetric signatures that can be related to various physical mechanisms 
that, from the radio band to the soft-$\gamma$ rays, probe increasingly smaller AGN regions. When we compile all the polarimetric data ever recorded for 
NGC~1068, by far the most observed radio-quiet type-2 AGN in polarization \cite{Marin2018}, we can take stock of our knowledge and the work that remains 
to be done.

\begin{figure}[t]
  \centering
  \includegraphics[scale=0.68]{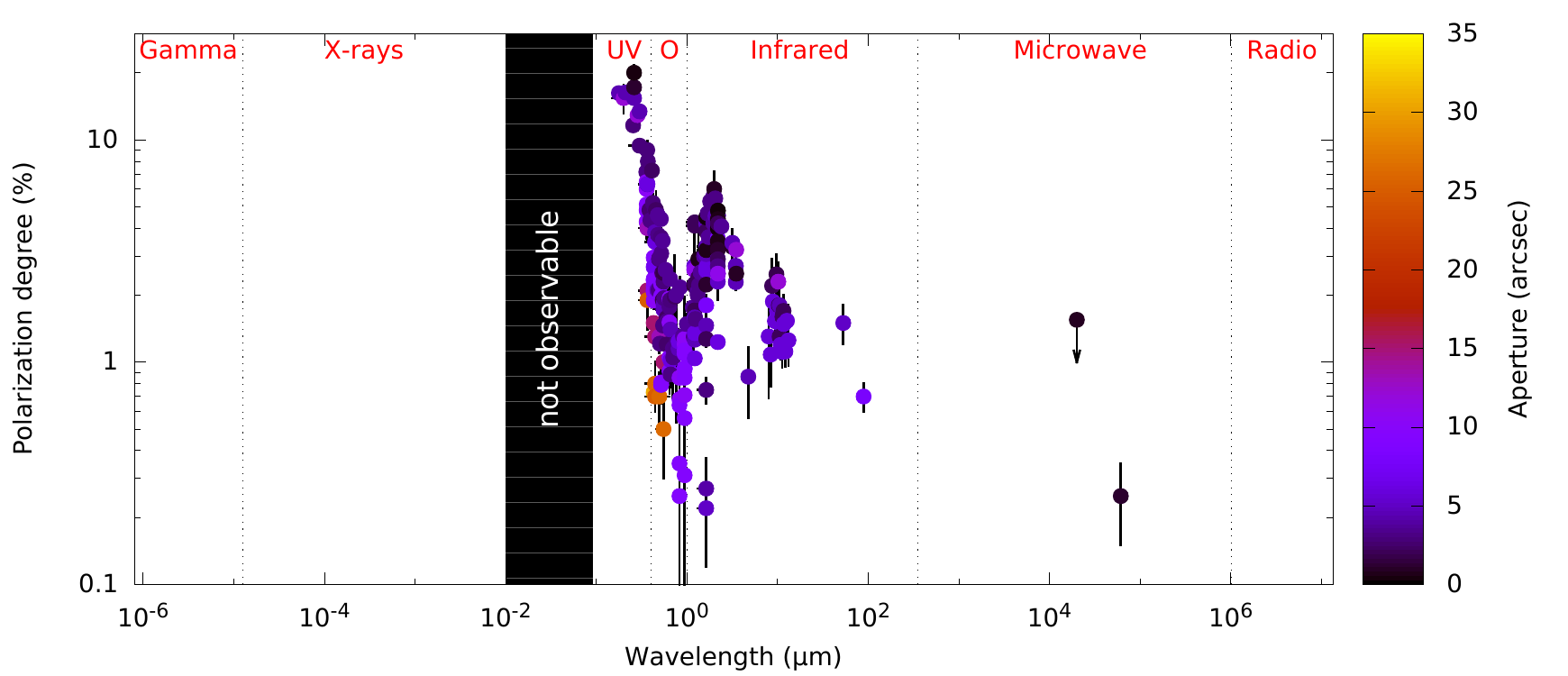}
  \caption{Summary of all the polarimetric measurements ever published for 
	    NGC~1068, by far the most observed radio-quiet type-2 AGN in 
	    polarization \cite{Marin2018}. Instrument apertures are 
	    color-coded (in arc-seconds) and the black area shows where
	    no measurements are achievable due to the hydrogen Galactic 
	    absorption downward 91.2~nm.}
  \label{Fig:Panchromatic_view}
\end{figure}

\begin{itemize}
\item{As it can be seen from Fig.~\ref{Fig:Panchromatic_view}, the $\gamma$-ray and X-ray bands are completely uncharted territories. The reasons is simple: 
in the past, there was never a high energy polarimeter sensitive enough to detect any polarization signature from AGNs. Not that the polarization 
degree is expected to be to low, but this was simply never tried for purely technical and economical reasons. This statement is true for all kind of AGNs, 
independently of the presence of jets. The high energy band is expected, in fact, to show detectable polarization levels \cite{McNamara2009,Marin2018a,Marin2018b}. 
A list of mechanisms that could produce $\gamma$ and X-ray polarization includes Compton and inverse-Compton scattering, synchrotron emission, 
Faraday rotation, bremsstrahlung, and collisionally excited line radiation from low energy cosmic protons, electrons and atoms \cite{Dolan1967}.}

\item{The extreme and far-ultraviolet bands are also lacking existing polarization measurements for similar reasons. In the optical band, polarization is 
observed using birefringent crystals. However, in the case of far-ultraviolet polarimetry, there is no birefringent material that can transmit light 
as crystals become opaques. A work around must be found using reflection polarimetry but the technique is yet to be optimized and tested \cite{Bolcar2017}. 
Scattering by electrons, atoms and dust grains, together with non-thermal emission are expected to provide strong polarization degrees since the diluting,
unpolarized starlight flux from the host galaxy strongly diminishes in the ultraviolet.}

\item{The mid- and near-ultraviolet, the optical, and the near-infrared polarization of AGNs has been recorded the most. White light polarization measurements
are achievable from ground telescopes, and the technology is mature. Multiple apertures and detectors have been used, leading to a wealth of polarimetric
data in those wavebands. The polarization is, however, strongly diluted by the host galaxy starlight, starburst light and interstellar polarization
(see the strong polarization deep in the optical band in Fig.~\ref{Fig:Panchromatic_view}). Removing the parasitic light is a complicated process that implies 
to observe both stars close to the AGN line-of-sight (to evaluate the importance of interstellar polarization) and the host galaxy. By minimizing the stellar 
absorption features in the residuals of the host/AGN flux ratio spectra, it is possible to estimate the starlight fractions in the observed AGN continuum flux. 
The situation can be even more complicated if synchrotron radiation from a jet superimposes its intrinsic polarization to the observation. So, even if white 
light polarimetry is common, it is not an easy task to interpret data.}

\item{The infrared band, loosely defined as 1 -- 300~$\mu$m here, has been less observed in polarimetric mode. Technologically speaking, it is more complex 
to observe infrared polarization since cryogenic cooling of the detectors in necessary in order to minimize the presence of thermal emission from warm 
optics. Below 2.5~$\mu$m, thermal emission from external hardwares is not too strong, so plarimeters can be constructed using half-wave plate retarders 
and birefringent crystals \cite{Jones1988}, explaining the presence of many measurements at 1 -- 2.2~$\mu$m in Fig.~\ref{Fig:Panchromatic_view}. In 
the mid- and far-infrared, there is a clear lack of polarimeters. Only a couple of points have been measured in the 19 -- 100~$\mu$m waveband using 
the SOFIA High-resolution Airborne Wideband Camera-plus \cite{Lopez2018,Marin2018}. Scattering, extinction and re-emission by aligned dust grains, together
with magnetic effects, are responsible for most of the polarization that one can detect in the infrared band of AGNs. }

\item{Finally, the millimeter and centimeter (alternatively called microwave and radio) polarization have been very little explored in the case of radio-quiet 
AGNs. When strong, kilo-parsec, collimated jets are present, they radiate a strong (polarized) flux. Non-thermal total and polarized synchrotron emission, 
differential Faraday rotation (with a $\lambda^2$ dependence), Faraday interconversion (from linear polarization to circular polarization, and vice-versa), 
and spectral depolarization (the polarization of a component changes sharply with frequency as the component becomes opaque) are responsible for the high
millimeter and centimeter polarization in radio-loud AGNs \cite{Leppanen1995}. In the case of radio-quiet AGNs, the source of radio emission and its 
associated polarization are much less known \cite{Laor2008,Ishibashi2011}. As an example, the degree of polarization at 4.9~GHz and 15~GHz measured for the 
radio-quiet AGN NGC~1068 (see Fig.~\ref{Fig:Panchromatic_view}) was at the sensitivity limits of the VLA observations. New polarimetric studies in the 
millimeter and centimeter bands are thus necessary.}
\end{itemize}

%%%%%%%%%%%%%%%%%%%%%%%%%%%%%%%%%%%%%%%%%%%%%%%%%%%%%%%%%%%%%%%%%%%%%%%%%%%%%%%%%%%%%%
%%%%%%%%%%%%%%%%%%%%%%%%%%%%%%%%%%%%%%%%%%%%%%%%%%%%%%%%%%%%%%%%%%%%%%%%%%%%%%%%%%%%%%
\section{Open questions that polarimetry could solve}
\label{Prospects}
Our knowledge of the panchromatic polarization of AGNs is fragmentary. In fact, looking at Fig.~\ref{Fig:Panchromatic_view}, only 24\% of the whole 
electromagnetic spectrum has been observed for radio-quiet AGNs. In the case of radio-loud objects, this percentage is about two times higher, but 
the high energy part of the spectrum is missing for both AGN types. New observations, both in the uncharted high energy band and in the other parts 
of the electromagnetic spectrum, are severely necessary. Numerous constraints on various physical scenarios and AGN hypotheses can be achieved with 
current and future polarimeters. I will review a few of them, roughly by increasing wavelength, with their associated future instruments.

\subsection{Probing the origin of the X-ray fluxes}
\label{Prospects:X_rays}
The morphology and physics of the region responsible for the emission of X-rays in AGNs is still highly debated. Almost all models rely on a compact 
corona situated above the accretion disk that is heated up by magnetic reconnection. The corona is believed to reprocess thermal ultraviolet photons 
emitted by the disk up to the X-rays by Inverse-Compton scattering, resulting in a corona that illuminates back the disk due to gravitational effects.
Distorted line emission around 6.4~keV indicates that the X-ray corona should be close to the disk so that strong gravity effects are dominant 
\cite{Merloni2003}. However, recent X-ray polarimetric observations by the PoGO+ pathfinder have shaken our certitudes about this scenario for the 
case of X-ray binaries in the hard state, i.e. when thermal emission from the disk dominates \cite{Chauvin2018}. Future X-ray polarimetric measurements 
with the Imaging X-ray Polarimetry Explorer (IXPE, \cite{Weisskopf2016}) or with the enhanced X-ray Timing and Polarimetry Mission (eXTP, \cite{Zhang2016})
will shed light on this question. The geometry of the corona can be probed thanks to the sensitivity of polarimetry to the geometrical shape of 
the emission source. A slab corona will produce a more asymmetric radiation pattern than a spherical one, and strong gravity effects will change the polarization
in a predictable way. The spin and the mass of the SMBH, together with the inclination of the accretion disk, will be within the reach of future 
X-ray polarimeters \cite{Schnittman2010,Dovciak2011}.

\subsection{Testing the accretion disk paradigm}
\label{Prospects:UV}
Some key signatures of accretion disks can be revealed only in polarized light (see Sect.~\ref{History:Disk_spectrum}), and with higher contrast at 
ultraviolet than at longer wavelengths. Specifically, models of disk atmospheres usually assume Compton scattering in an electron-filled plasma, 
resulting in inclination-dependent polarization signatures. We would expect a polarization degree in the range 0 to 11.7\%, and always lying along 
the (projected) disk plane \cite{Chandrasekhar1960}. Yet optical polarization in type-1 objects is almost always detected at less than a percent
and parallel to the radio jets if any (so perpendicular to our expectations) \cite{Stockman1979,Antonucci2002}. Whether these low levels 
can be attributed to dominant absorption opacity \cite{Laor1989} or complete Faraday depolarization \cite{Agol1996} is unclear. This degeneracy 
may be broken by looking at the numerous ultraviolet lines that are formed in the innermost AGN regions (e.g, Ly$\alpha~\lambda$1216, 
C{\sc ii}$~\lambda$1335, C{\sc iv}$~\lambda$1549, Mg{\sc ii}$~\lambda$2800). These lines are the key to test the accretion disk paradigm
and only ultraviolet polarimetric observations with high signal-to-noise ratio and high spectral resolution can distinguish between the two effects. 
In this context, a high resolution spectropolarimeter such as POLLUX can provide the most exquisite results. POLLUX is an instrument envisioned for 
the 15-meter primary mirror option of LUVOIR (a multi-wavelength space observatory concept being developed by the Goddard Space Flight Center and 
proposed for the 2020 Decadal Survey Concept Study \cite{Bouret2018}). The ultraviolet coverage and large resolution of POLLUX would also allow to
test models of accretion disk atmospheres, where the ionization of hydrogen should give rise to spectral features at the Lyman limit (912~\AA). Yet,
the Lyman edge feature has never been convincingly observed in total flux \cite{Kriss1999,Shull2012}. Polarized flux observations could remove the
parasitic light from the other AGN components and allow to better detect the presence of a sharp discontinuity in the continuum at the Lyman limit 
\cite{Kishimoto2005}.

\subsection{Revealing the location, composition and geometric arrangement of dust}
\label{Prospects:IR}
Decomposition of the SED of AGNs suggests that the mid-infrared component corresponds to equatorial emission, approximately aligned with the plane 
of the inner accretion disk, while the weaker near-infrared peak might be associated to hot dust in the inner polar region. However, recent studies 
revealed that this picture is probably erroneous \cite{Honig2013,Asmus2016}. The bulk of the infrared emission seems to originate from the polar 
region above the circumnuclear dust, where only little dust should be present. Using new high angular resolution polarimetric observations with 
adaptive optics systems \cite{Gratadour2015,Lopez2016}, it will be possible to investigate the true location of dust in AGNs and derive its 
polarization-sensitive composition. Moreover, as polarization is enhanced by paramagnetic dust grains alignment, it is possible to go beyond the 
capabilities of interferometry and probe the parsec-scale magnetic fields. The key argument here is that the magnetic fields align the dust grains 
according to their intensity and direction. By fitting the infrared polarization data with different numerical models, the topology of parsec scales 
magnetic fields can be constrained \cite{Peest2017,Grosset2018}. Understanding the observed gradual rotation of the polarization angle towards the 
far-infrared in NGC~1068 or 3C~273 can allow a characterization of the global coherent magnetic field structure impacting dust scattering, 
absorption, and emission over the full infrared band emission.

\subsection{The physics and the internal structure of the innermost jet regions}
\label{Prospects:Radio}
AGNs routinely accelerate significant fractions of a solar mass to near light speeds in the form of ballistic jets. Due to the ordered magnetic fields
and directionality of emission, the emitted synchrotron radiation is naturally highly polarized \cite{Zheleznyakov2002}. Blazars, a sub-class of AGNs that 
have the orientation of their jets close to the line of sight, are particularly interesting on this case since their orientation makes their non-thermal
radiation highly relativistically beamed, and thus very bright. The combination of brightness and high polarization allows new polarimetric imaging in 
the (sub)millimeter and radio bands by the Atacama Large Millimeter/submillimeter Array (ALMA) to identify magnetic field configurations at unprecedented 
scales \cite{Liuzzo2015}. Past NRAO telescopes (VLA, GBT, VLBA) have proven to be particularly useful by, e.g., revealing a rotation of the position angle 
of linear polarization in the jet of BL~Lacertae, following a compressive feature propagating down the helical jet field \cite{Marscher2008}. However, 
the case is more complex for radio-quiet AGN where the expected polarized radio flux is too low compared to the sensitivity of old radio telescopes. 
There is essentially no exploitable data and only future observations of radio-quiet AGNs, using the most recent radio observatories, will provide polarimetric 
information at millimeter and centimeter wavelengths. Detecting the radio polarization signal from radio-quiet AGNs is fundamental to shed light on 
the physical mechanisms producing the faint radio emission. Looking at high angular resolution polarization maps of jet's bases will also allow to better
constrain magneto-hydrodynamic and plasma models \cite{Agudo2018}, especially through the prism of the enigmatic variability processes.

%%%%%%%%%%%%%%%%%%%%%%%%%%%%%%%%%%%%%%%%%%%%%%%%%%%%%%%%%%%%%%%%%%%%%%%%%%%%%%%%%%%%%%
%%%%%%%%%%%%%%%%%%%%%%%%%%%%%%%%%%%%%%%%%%%%%%%%%%%%%%%%%%%%%%%%%%%%%%%%%%%%%%%%%%%%%%
\section{Summary}
\label{Conclusions}
Panchromatic polarization measurements of AGNs are necessary to better understand how such objects can form, accrete matter and re-emit copious 
amounts of particles and radiation, ultimately impacting the host galaxy they reside in. In the past, polarimetry has brought the most important 
constraints on the true nature of AGNs, and it keeps revealing more and more complex features associated with accretion, magnetic, emission,
absorption and re-emission processes. Polarization, at least in the case of quasi-stellar radio sources, is complex to decipher due to the 
presence of a contaminating flux from host galaxy. Yet, one can access fundamental signatures of spatially unresolvable AGN components thanks 
to polarization. Many questions linked with AGN formation and evolution processes remain open, and only future polarimetric measurements using 
30-meters class telescopes or satellites equipped with state-of-the-art polarimeters can resolve them. In particular, accretion disk theories, 
disk atmosphere models and jet production mechanisms will be within our reach in the next decade thanks to the renewed enthusiasm of the community 
about high energy polarimetry \cite{Marin2018c}. The construction of large radio and millimeter telescopes will also fully benefit the field by 
probing sub-arcsecond scales. The visible and infrared bands are maybe lagging behind in new instrument projects, yet those wavebands are equally 
important as the other parts of the electromagnetic spectrum for a complete understanding of AGNs.

\begin{acknowledgement}
I would like to deeply thank Damien Hustem\'ekers, Delphine Porquet and Robert ``Ski'' Antonucci for their insightful comments that helped to 
improve the quality of this chapter.
\end{acknowledgement}

\input{referenc}

\end{document}

%% file: referenc.tex
%%%%%%%%%%%%%%%%%%%%%%%% referenc.tex %%%%%%%%%%%%%%%%%%%%%%%%%%%%%%
% sample references
% %
% Use this file as a template for your own input.
%
%%%%%%%%%%%%%%%%%%%%%%%% Springer-Verlag %%%%%%%%%%%%%%%%%%%%%%%%%%
%
% BibTeX users please use
% \bibliographystyle{}
% \bibliography{}
%